\documentclass[12pt]{article}
\usepackage{a4,amsmath,cite,amssymb,graphicx,epsfig}

\def\ecart{\noalign{\medskip}}

\begin{document}
\begin{titlepage}

\vspace*{1cm}

\begin{center} 
\setlength{\baselineskip}{24pt}
{\LARGE HOW CAN THE ODDERON BE DETECTED  \\
AT RHIC AND LHC}
\end{center}

\begin{center}
\vspace{2cm}
{\large Regina F. Avila, Pierre Gauron, Basarab Nicolescu}

\vspace{0.5cm} 
 Theory Group,  Laboratoire de Physique Nucl\'eaire  et des Hautes  \'Energies 
(LPNHE)\footnote{Unit\'e  de Recherche des Universit\'es 
  Paris 6 et Paris 7, Associ\'ee au CNRS},
 CNRS and Universit\'e Pierre et Marie Curie, Paris\\ 

\end{center}

\vspace{2cm}

\textbf{Abstract}: 
The Odderon remains an elusive object, 33 years after its invention. 
The Odderon is now a fundamental object in QCD and CGC and it has to be found experimentally if QCD and CGC are right.  
In the present paper, we show how to find it at RHIC and LHC. 
The most spectacular signature of the Odderon is the predicted difference between the differential cross-sections for proton-proton and antiproton-proton at high $s$ and moderate $t$. 
This experiment can be done by using the STAR detector at RHIC and by combining these future data with the already present UA4/2 data.  
The Odderon could also be found by ATLAS experiment at LHC by performing a high-precision measurement of the real part of the hadron elastic scattering amplitude at small $t$.

\end{titlepage}
\newpage

%************************************************
\section{Introduction}
%************************************************
The Odderon is defined as a singularity in the complex J-plane, located at $J=1$ when $t=0$ and which contributes to the odd-under-crossing amplitude $F_-$.
The concept of Odderon first emerged in 1973 in the context of asymptotic theorems~\cite{Lukaszuk:1973nt}.
7 years later, it was possibly connected with 3-gluon exchanges in perturbative QCD~\cite{Bartels:1980pe,Jaroszewicz:1980mq,Kwiecinski:1980wb}, but it took 27 years to firmly rediscover it in the context of pQCD~\cite{Bartels:1999yt}.
The Odderon was also rediscovered recently in the Color Glass Condensate (CGC) approach~\cite{Hatta:2005as,Jeon:2005cf} and in the dipole picture~\cite{Kovchegov:2003dm}.
One can therefore assert that the Odderon is a crucial test of QCD.

On experimental level, there is a strong evidence for the non-perturbative Odderon: the discovery, in 1985, of a difference between $(d\sigma/dt)_{\bar pp}$ and 
$(d\sigma/dt)_{pp}$ in the dip-shoulder region $1.1<\vert t\vert <1.5$ GeV$^2$ at $\sqrt{s}=$ 52.8 GeV~\cite{Breakstone:1985pe,Erhan:1984mv}.
Unfortunately, these data were obtained in one week, just before ISR was closed and therefore the evidence, even if it is strong (99,9 \% confidence level), is not totally convincing.
Moderate evidence for the existence of the non-perturbative Odderon also comes from the dramatic change of shape in the polarization in $\pi^-p\to\pi^0n$, in going form $p_L=5$ GeV/c~\cite{Hill:1973bq,Bonamy:1973dz} to $p_L=40$~GeV/c~\cite{Apokin:1981mb}, but this Odderon corresponds to a different type of Odderon as compared with the one identified in pQCD.
Finally, weak evidence for the non-perturbative Odderon comes from a strange structure seen in the UA4/2 $dN/dt$ data for $\bar pp$ scattering at $\sqrt{s}=541$~GeV, namely a bump centered at $\vert t\vert=2\cdot 10^{-3}$ GeV$^2$~\cite{Augier:1993sz}.
This structure could correspond to oscillations of a very small period due to the presence of the Odderon~\cite{Gauron:1996sm}.

All the above mentioned experimental results point towards the maximal Odderon~\cite{Lukaszuk:1973nt,Nicolescu:2005bn}, a special case corresponding to the maximal asymptotic $(s\to\infty)$ behavior allowed by the general principles of strong interactions:
\begin{equation}
\label{eq:1}
\sigma_T(s)\propto \ln^2s,\quad\mbox{as } s\to\infty
\end{equation}
and
\begin{equation}
\label{eq:2}
\Delta\sigma(s)\equiv \sigma_T^{\bar pp}(s)-\sigma_T^{pp}(s)\propto \ln s,\quad\mbox{as } s\to\infty\ .
\end{equation}
Interestingly enough, an important stream of theoretical papers concern precisely the maximal behavior~\cite{Lukaszuk:1973nt}, which was first discovered by Heisenberg in 1952~\cite{Heisenberg:1952zp} and later proved, in a more rigorous way by Froissart and Martin~\cite{Froissart:1961ux,Martin:1965jj}.
Half a century after the discovery of Heisenberg, this maximal behavior~(\ref{eq:1}) was also proved in the context of the AdS/CFT dual string-gravity theory~\cite{Giddings:2002cd} and of the Color Glass Condensate approach~\cite{Ferreiro:2002kv}.
It was also shown to provide the best description of the present experimental data on total cross-sections~\cite{Cudell:2001pn,Eidelman:2004wy}.

Of course, the experimental indication of the maximal behavior~(\ref{eq:1}) is not \textit{per se} an indication for the maximal Odderon behavior~(\ref{eq:2}): the imaginary part of the even-under-crossing amplitude $F_+$ can very well behave like $\ln^2s$ for $s\to\infty$ and, at the same time, the imaginary part of the odd-under crossing amplitude $F_-$ can vanish for $s\to\infty$.
But this would be a very unnatural situation: the maximal behavior of $ImF_+(s,t=0)\propto\ln^2s$ is naturally associated with the maximal behavior $ImF_-(s,t=0)\propto\ln s$.
In other words, strong interactions should be as strong as possible.

In the present paper we will consider a very general form of the hadron amplitudes compatible with both the maximal behavior of strong interaction at asymptotic energies and with the well established Regge behavior at moderate energies, i.e. at pre-ISR and ISR energies~\cite{Gauron:1986nk,Gauron:1989cs}.

Our strategy is the following:
\begin{itemize}
  \item [1.]
  We will consider two cases: one in which the Odderon is absent and one in which the Odderon is present.
  \item [2.]
  We will use the two respective forms in order to describe the 832 experimental points for $pp$ and $\bar pp$ scattering, from PDG Tables, for $\sigma_T(s),\ \rho(s)$ and 
  $d\sigma/dt(s,t)$, in the s-range
  \begin{equation}
\label{eq:3}
4.539\mbox{ GeV }\leqslant\sqrt{s}\leqslant 1800\mbox{ GeV}
\end{equation}
and in the t-range
\begin{equation}
\label{eq:4}
0\leqslant \vert t\vert \leqslant 2.6\mbox{ GeV}^2\ .
\end{equation}
The best form will be chosen.
  \item [3.]
  In order to make predictions at RHIC and LHC energies, we will insist on the best possible \textit{quantitative} description of the data.
  Most of the existing phenomenological  models describe only the gross features of the data in a limited region of energy and therefore they could lead to wrong quantitative predictions at much higher energies, especially in looking for such delicate effects like those associated with the presence of the Odderon.
  Models which describe gross features of the existing data, can, at best, describe gross features of future data at RHIC and LHC energies.
 \item [4.] 
 From the study of the interference between $F_+(s,t)$ and  $F_-(s,t)$ amplitudes we will conclude which are the best experiments to be done in order to detect in a clear way the Odderon.
\end{itemize}

%************************************************
\section{The form of the amplitudes}
%************************************************
$F_\pm$ are defined to be
\begin{equation}
\label{eq:5}
F_\pm(s,t)=\frac{1}{2}\left(F_{pp}(s,t)\pm F_{\bar pp}(s,t)\right)
\end{equation}
and are normalized so that
\begin{equation}
\label{eq:6}
\sigma_T(s)=\frac{1}{s}\mathrm{Im} F(s,0)\ ,\quad
\rho(s)=\frac{\mathrm{Re}F(s,t=0)}{\mathrm{Im}F(s,t=0)}
\end{equation}
\begin{equation}
\label{eq:7}
\frac{d\sigma}{dt}(s,t)=\frac{1}{16\pi s^2}\vert F(s,t)\vert^2\ .
\end{equation}
The $F_+(s,t)$ amplitude is written as a sum of the following components~\cite{Gauron:1986nk,Gauron:1989cs}:
\begin{itemize}
  \item [a)]
  $F_+^H(s,t)$ representing the contribution of a 3/2 - cut collapsing, at $t=0$, to a triple pole located at $J=1$ and which satisfies the Auberson-Kinoshita-Martin asymptotic theorem~\cite{Auberson:1971ru}:
\begin{equation}
\label{eq:8}
\begin{array}{rcl}
     \dfrac{1}{is} F_+^{H}(s,t) & = & H_1\ln^2\bar s\ \frac{2J_1(K_+\bar\tau)}{K_+\bar\tau}
\exp(b_1^+t) \\
\ecart
      &   + & H_2\ln\bar sJ_0(K_+\bar\tau)\exp(b_2^+t)\\
\ecart
      &   + & H_3[J_0(K_+\bar\tau)-K_+\bar\tau J_1(K_+\bar\tau)]\exp(b_3^+t)\ ,
\end{array}
\end{equation}  
where $J_n$ are Bessel functions, $H_k,\ b_k^+(k=1,2,3)$ and $K_+$ are constants,
\begin{equation}
\label{eq:9}
\bar s=\left(\frac{s}{s_0}\right)\exp\left(-\frac{1}{2}i\pi\right),
\mbox{ with }s_0=1\mbox{ GeV}^2	
\end{equation}
and
\begin{equation}
\label{eq:10}
\bar\tau=\left(-\frac{t}{t_0}\right)^{1/2}\ln\bar s,\quad \mbox{with }t_0=1\mbox{ GeV}^2\ .
\end{equation}
Let us note that, by putting $t=0$ in eq.~(\ref{eq:8}), we get from eq.~(\ref{eq:6}) that $\sigma_T(s)$ has a component expressed, at \textit{finite} energies, as a quadratic form in $\ln s$, i.e. precisely the form discovered by Heisenberg~\cite{Heisenberg:1952zp}.
This justifies the index $H$ in eq.~(\ref{eq:8}).
  \item [b)]
  $F_+^P(s,t)$, the contribution of the Pomeron Regge pole:
\begin{equation}
\label{eq:11}
\dfrac{1}{s} F_+^P(s,t)=C_P\exp(\beta_Pt)[i-\cot(\frac{\pi}{2}\alpha_P(t))]\left(\dfrac{s}{s_0}\right)^{\alpha_P(t)-1}\ ,
\end{equation}
where $C_P$ and $\beta_P$ are constants, and
\begin{equation}
\label{eq:12}
\alpha_P(t)=\alpha_P(0)+\alpha'_Pt\ ,
\end{equation}
with
\begin{equation}
\label{eq:13}
\alpha_P(0)=1 \quad \mbox{ and }\quad \alpha'_P=0.25\mbox{ GeV}^{-2}\ .
\end{equation}
  \item [c)]
 $F_+^{PP}(s,t)$ , the contribution of the Pomeron-Pomeron Regge cut:
\begin{multline}
\label{eq:14}
\dfrac{1}{s} F_+^{PP}(s,t)=C_{PP}\exp(\beta_{PP}t)
[i\sin(\frac{\pi}{2}\alpha_{PP}(t))-\cos(\frac{\pi}{2}\alpha_{PP}(t))]\\
\times\frac{(s/s_0)^{\alpha_{PP}(t)-1}}{\ln[(s/s_0)\exp(-\frac{1}{2}i\pi)]}\ ,
\end{multline}
where $C_{PP}$ and $\beta_{PP}$ are constants, and
\begin{equation}
\label{eq:15}
\alpha_{PP}(t)=\alpha_{PP}(0)+\alpha'_{PP}t\ ,
\end{equation}
with 
\begin{equation}
\label{eq:16}
\alpha_{PP}(0)=1 \quad \mbox{ and }\quad \alpha'_{PP}=\frac{1}{2}\alpha'_P\ .
\end{equation}
  \item [d)]
$F_+^R(s,t)$, the contribution of a secondary Regge trajectory, whose intercept is located around $J=\frac{1}{2}$ and associated with the $f_0(980)$ and $a_0(980)$ particles:
\begin{equation}
\label{eq:17}
\dfrac{1}{s} F_+^R(s,t)=C_R^+\gamma_R^+(t)\exp(\beta_R^+t)
[i-\cot(\frac{1}{2}\pi\alpha_R^+(t))]\left(\dfrac{s}{s_0}\right)^{\alpha_R^+(t)-1}\ ,
\end{equation} 
where
\begin{equation}
\label{eq:18}
\alpha_R^+(t)=\alpha_R^+(0)+(\alpha'_R)^+t\ ,
\end{equation} 
with $(\alpha'_R)^+$ fixed at the world phenomenological value 0.88 GeV$^{-2}$, and
\begin{equation}
\label{eq:19}
\gamma_R^+(t)=\frac{\alpha_R^+(t)[\alpha_R^+(t)+1][\alpha_R^+(t)+2]}
{\alpha_R^+(0)[\alpha_R^+(0)+1][\alpha_R^+(0)+2]}\ ,
\end{equation}
$C_R^+,\ \beta_R^+$ and $\alpha_R^+(0)$ being constants;
	\item [e)]
$F_+^{RP}(s,t)$, the contribution of the reggeon-Pomeron Regge cut:
\begin{multline}
\label{eq:20}
\dfrac{1}{s} F_+^{RP}(s,t)=\left(\dfrac{t}{t_0}\right)^2C_{RP}^+\exp(\beta_{RP}^+t)
[i\sin(\frac{\pi}{2}\alpha_{RP}^+(t))-\cos(\frac{\pi}{2}\alpha_{RP}^+(t))]\\
\times\frac{(s/s_0)^{\alpha_{RP}^+(t)-1}}{\ln[(s/s_0)\exp(-\frac{1}{2}i\pi)]}\ ,
\end{multline}
\begin{equation}
\label{eq:21}
\alpha_{RP}^+(t)=\alpha_{RP}^+(0)+(\alpha'_{RP})^+t\ ,
\end{equation}
where $C_{RP}^+,\ \beta_{RP}^+$ and $\alpha_{RP}^+(0)$ are constants, and
\begin{equation}
\label{eq:22}
(\alpha_{RP}^{\prime})^+=\frac{(\alpha_{R}^{\prime})^+\alpha_{P}^{\prime}}
{(\alpha_{R}^{\prime})^++\alpha_{P}^{\prime}}\ .
\end{equation}
\end{itemize}

The even-under-crossing amplitude $F_+(s,t)$ is therefore defined as:
\begin{equation}
\label{eq:23}
F_+(s,t)=F_+^{H}(s,t)+F_+^P(s,t)+F_+^{PP}(s,t)+F_+^R(s,t)+F_+^{RP}(s,t)\ .
\end{equation}

In its turn, the $F_-(s,t)$ amplitude is written as a sum of the following components~\cite{Gauron:1986nk,Gauron:1989cs}:
\begin{itemize}
  \item [a)]
  $F_-^{MO}(s,t)$ representing the maximal Odderon contribution, resulting from two complex conjugate poles collapsing, at $t=0$, to a dipole located at $J=1$ and which satisfies the Auberson-Kinoshita-Martin asymptotic theorem:
\begin{equation}
\label{eq:24}
  \frac{1}{s}F_-^{MO}(s,t)=O_1\ln^2\bar s\frac{\sin(K_-\bar\tau)}{K_-\bar\tau}\exp(b_1^-t)+
O_2\ln\bar s\cos(K_-\bar\tau)\exp(b_2^-t)+O_3\exp(b_3^-t)\ ,
\end{equation}
where $O_k,\ b_k^-(k=1,2,3)$ and $K_-$ are constants.
  \item [b)]
  $F_-^O(s,t)$, the contribution of the minimal Odderon Regge pole:
\begin{equation}
\label{eq:25}
\dfrac{1}{s} F_-^O(s,t)=C_O\exp(\beta_Ot)
[i+\tan(\frac{1}{2}\pi\alpha_O(t))]\left(\dfrac{s}{s_0}\right)^{\alpha_O(t)-1}
[1+\alpha_O(t)][1-\alpha_O(t)]\ ,
\end{equation}  
where $C_O$ and $\beta_O$ are constants and
\begin{equation}
\label{eq:26}
\alpha_O(t)=\alpha_O(0)+\alpha'_Ot\ ,
\end{equation}
with
\begin{equation}
\label{eq:27}
\alpha_O(0)=1\ .
\end{equation}
  \item [c)]
  $F_-^{OP}(s,t)$, the contribution of the minimal Odderon-Pomeron Regge cut:
\begin{multline}
\label{eq:28}
\dfrac{1}{s} F_-^{OP}(s,t)=C_{OP}\exp(\beta_{OP}t)
[\sin(\frac{1}{2}\pi\alpha_{OP}(t))+i\cos(\frac{1}{2}\pi\alpha_{OP}(t))]\\
\times\frac{(s/s_0)^{\alpha_{OP}(t)-1}}{\ln[(s/s_0)\exp(-\frac{1}{2}i\pi)]}\ ,
\end{multline} 
where $C_{OP}$ and $\beta_{OP}$ are constants, and
\begin{equation}
\label{eq:29}
\alpha_{OP}(t)=\alpha_{OP}(0)+\alpha'_{OP}t\ ,
\end{equation} 
with
\begin{equation}
\label{eq:30}
\alpha_{OP}(0)=1
\end{equation}
and
\begin{equation}
\label{eq:31}
\alpha'_{OP}=\frac{\alpha'_O\cdot\alpha'_P}{\alpha'_O+\alpha'_P}
\end{equation}
\item [d)]
$F_-^R(s,t)$, the contribution of a secondary Regge trajectory located around $J=1/2$ and associated with the $\rho(770)$ and $\omega(782)$ particles:
\begin{equation}
\label{eq:32}
\dfrac{1}{s} F_-^R(s,t)=-C_R^-\gamma_R^-(t)\exp(\beta_R^-t)
[i+\tan(\frac{1}{2}\pi\alpha_R^-(t))]\left(\dfrac{s}{s_0}\right)^{\alpha_R^-(t)-1}\ ,
\end{equation}
\begin{equation}
\label{eq:33}
\alpha_R^-(t)=\alpha_R^-(0)+(\alpha'_R)^-t\ ,
\end{equation}

\vspace*{0.5cm}
with $(\alpha'_R)^-$ fixed at the value 0.88 GeV$^2$, and
\begin{equation}
\label{eq:34}
\gamma_R^-(s,t)=\frac{\alpha_R^-(t)[\alpha_R^-(t)+1][\alpha_R^-(t)+2]}
{\alpha_R^-(0)[\alpha_R^-(0)+1][\alpha_R^-(0)+2]}\ ,
\end{equation}
$C_R^-,\ \beta_R^-$ and $\alpha_R^-(0)$ being constants.
\item [e)]
$F_-^{RP}(s,t)$, the contribution of the reggeon-Pomeron Regge cut:
\begin{multline}
\label{eq:35}
\dfrac{1}{s} F_-^{RP}(s,t)=\left(\dfrac{t}{t_0}\right)^2C_{RP}^-\exp(\beta_{RP}^-t)
[\sin(\frac{\pi}{2}\alpha_{RP}^-(t))+i\cos(\frac{\pi}{2}\alpha_{RP}^-(t))]\\
\times\frac{(s/s_0)^{\alpha_{RP}^-(t)-1}}{\ln[(s/s_0)\exp(-\frac{1}{2}i\pi)]}\ ,
\end{multline}
\begin{equation}
\label{eq:36}
\alpha_{RP}^-(t)=\alpha_{RP}^-(0)+(\alpha'_{RP})^-t\ ,
\end{equation}
where $C_{RP}^-,\ \beta_{RP}^-$ and $\alpha_{RP}^-(0)$ are constants and
\begin{equation}
\label{eq:37}
(\alpha_{RP}^{\prime})^-=\frac{(\alpha_{R}^{\prime})^-\alpha_{P}^{\prime}}
{(\alpha_{R}^{\prime})^-+\alpha_{P}^{\prime}}\ .
\end{equation}
\end{itemize}

The odd-under-crossing amplitude $F_-(s,t)$ is there defined as:
\begin{equation}
\label{eq:38}
F_-(s,t)=F_-^{MO}(s,t)+F_-^O(s,t)+F_-^{OP}(s,t)+F_-^R(s,t)+F_-^{RP}(s,t)\ .
\end{equation}
We can now write the $pp$ and $\bar pp$ amplitudes as:
\begin{equation}
\label{eq:39}
F_{pp}(s,t)=F_+(s,t)+F_-(s,t)
\end{equation}
and
\begin{equation}
\label{eq:40}
F_{\bar pp}(s,t)=F_+(s,t)-F_-(s,t)
\end{equation}
where $F_+(s,t)$ and $F_-(s,t)$ are defined through eqs.~(\ref{eq:8})-(\ref{eq:23}) and (\ref{eq:24})-(\ref{eq:40}), respectively.
The observables $\sigma_T(s),\ \rho(s,t)$ and $(d\sigma/dt)$ are evaluated through eqs.~(\ref{eq:39}), (\ref{eq:40}), (\ref{eq:6}) and (\ref{eq:7}).
%************************************************
\section{Numerical results}
%************************************************
\subsection{The case without the Odderon}
%************************************************

Let us first consider the case \textit{without the Odderon}, i.e. the case with
\begin{equation}
\label{eq:41}
O_k=0\ (k=1,2,3),\quad C_O=0,\quad C_{OP}=0.
\end{equation}
In this case, one has 23 free parameters:
$H_k,\ b_k^+\ (k=1,2,3),\ K_+,\ C_P,\ \beta_P,\ C_{PP},\ \beta_{PP},\ C_R^+,$
$\beta_R^+,\ \alpha_R^+(0),\ C_{RP}^+,\ \beta_{RP}^+, 
\alpha_{RP}^+(0),\ C_R^-,\ \beta_R^-,\ \alpha_R^-(0),\ C_{RP}^-,\ \beta_{RP}^-$ and $\alpha_{RP}^-(0)$. 
The best values of these free parameters are obtained through a $\chi^2$ MINUIT minimization.

In spite of the quite impressive number of free parameters, the $\chi^2$-value is inacceptably bad:
\begin{equation}
\label{eq:42}
\chi^2/dof=14.2\ .
\end{equation}
A closer examination of the results reveals however an interesting fact: the no-Odderon case describes nicely the data in the t-region
\begin{equation}
\label{eq:43}
0\leqslant\vert t\vert\leqslant 0.6 \mbox{ GeV}^2
\end{equation}
but totally fails to describe the data for higher t-values, as exemplified in Fig.~1, where we represent the $pp$ and $\bar pp$ $d\sigma/dt$ data at $\sqrt{s}=52.8$ GeV together with the theoretical description of the no-Odderon case.

This failure is not due to the absence of the odd-under-crossing amplitude, because, even if the Odderon contributions are absent, we still have the pole and cut contributions, $F_-^R(s,t)$ and $F_-^{RP}(s,t)$.
However, these contributions fail to interfere with the $F_+(s,t)$ contributions in a correct way, for two physically important reasons:
\begin{itemize}
  \item [1)]
  The fact that the intercept of the trajectory of the secondary reggeon of odd signature is  half a unit lower that the Pomeron pole intercept induces a fast decrease with the energy of the secondary reggeon contributions and therefore the near equality of $pp$ and $\bar pp\ d\sigma/dt$, in contrast with the data.
  \item [2)]
  The fact that the past phenomenology imposes universal numerical values of the slopes $(\alpha'_R)^+,\ (\alpha'_R)^-$ and $\alpha'_P$ (they \textit{are not} free parameters) induces a decrease of $d\sigma/dt$ in $t$ at fixed $s$ and sufficiently high $t$, which is faster than what 
  $d\sigma/dt$ data indicate at ISR at CERN collider energies.
  In particular, the moderate t-region in the UA4/2 $d\sigma/dt$ is very badly described.
 \end{itemize}
 The failure of the above considered amplitudes to describe the data in the moderate t-region, does not mean the failure of the Regge model, which is a basic ingredient of the approach presented in this paper.
 It simply means the need for the Odderon.
 The Regge pole model is justified not only by the existing data at moderate $t\leqslant 0$ but also by the multitude of the resonances present in \textit{Review of Particle Physics}~\cite{Eidelman:2004wy}, which constitute a striking evidence for linear Regge trajectories with universal slope.
 The Regge pole model has to be included as a basic ingredient in any more sophistical approach aiming to a realistic description of the experimental data.
 It is very encouraging that pQCD already get Regge behavior ; in particular, gluons are reggeized in pQCD.
%************************************************
\subsection{The case with the Odderon}
%************************************************
In this case we have 12 supplementary free parameters as compared with the no-Odderon case:
$O_k,\ b_k^-\ (k=1,2,3),\ K_-,\ C_O,\ \beta_O,\ \alpha'_O,\ C_{OP}$ and $\beta_{OP}$.
 
The total of 35 free parameters of our approach could be considered, at a superficial glance, as too big.
However, one has to realize that the 23 free parameters associated with the dominant $F_+(s,t)$ amplitude and with the component of $F_-(s,t)$ responsible for describing the data for $\Delta\sigma(s)$ (see eq.~(\ref{eq:2})) and $\Delta\rho(s,t=0)$, where
\begin{equation}
\label{eq:44}
\Delta\rho(s,t=0)\equiv\rho^{\bar pp}(s,t=0)-\rho^{pp}(s,t=0)
\end{equation}
are, almost all of them, well constrained.

Moreover, the discrepancy between he no-Odderon model and the experimental data in the moderate-t region (especially at $\sqrt{s}=52.8$ GeV and $\sqrt{s}=541$ GeV) is so big that, in their turn, the supplementary 12 free parameters (at least, most of them) are also well constrained.

Let us also note that the above - mentioned discrepancy in the region of $t$ defined by
\begin{equation}
\label{eq:45}
0.6< \vert t\vert \leqslant 2.6 \mbox{ GeV}^2
\end{equation}
can not come, as one could thing, from the contributions induced by  perturbative QCD.
The region (\ref{eq:45}) is fully in the domain of validity of the non-perturbative Regge pole model and the respective values of $t$ are too small in order to make pQCD calculations.

The best values of the free parameters are obtained by again performing a $\chi^2$ MINUIT minimization.
Their numerical values are shown in Table 1.

%*****************
%******************
\begin{table}[h]
  \centering
\begin{tabular}{ c c c c c c c  }
\hline
  \multicolumn{3}{l}{Parameters of $F_+^H(s,t)$} &  &  &  &     \\
  $H_1$ & $b_1^+$ & $H_2$ & $b_2^+$ & $H_3$ & $b_3^+$ & $K_+$ \\
  (mb) & (GeV$^{-2})$ &  (mb)  & (GeV$^{-2})$ &  (mb)  & (GeV$^{-2})$ &  \\ 
\hline
0.4030 & 4.5691 & -3.8616 & 7.1798 & 9.2079  & 6.0270 & 0.6571\\
$\pm$ 0.0015 & $\pm$ 0.0677 & $\pm$ 0.0262 & $\pm$ 0.1603 & $\pm$ 0.2091 &
$\pm$ 0.0808 & $\pm$ 0.0089\\
\hline
\ecart
\ecart
\hline
 \multicolumn{3}{l}{Maximal Odderon parameters} &  &  &  &     \\
  $O_1$ & $b_1^-$ & $O_2$ & $b_2^-$ & $O_3$ & $b_3^-$ & $K_-$ \\
  (mb) & (GeV$^{-2})$ &  (mb)  & (GeV$^{-2})$ &  (mb)  & (GeV$^{-2})$ &  \\ 
\hline
-0.0690 & 8.9526 & 1.4166 & 3.4515 & -0.3558  & 1.1064 & 0.1267 \\
$\pm$ 0.0043 & $\pm$ 1.6989 & $\pm$ 0.0324 & $\pm$ 0.0361 & $\pm$ 0.0097 &
$\pm$ 0.0186 & $\pm$ 0.0017\\
\hline
\end{tabular}

\vspace{1cm}
\begin{tabular}{c | c c c c c c c c  }
\hline
\multicolumn{4}{l}{Reggeon poles and cuts parameters} &  &  &  &  &   \\
& P & PP & O & OP & $R_+$ & $R_-$ & $(RP)_+$ & $(RP)_-$ \\
$\alpha(0)$ & & & & &  0.48            & 0.34            & -0.56           & 0.70 \\
                   & & & & &  $\pm$ 0.01 & $\pm$ 0.02 & $\pm$ 0.06 & $\pm$ 0.20 \\
\hline
C                & 40.43 & -9.20 & -6.07 & 11.83 & 38.18 & 47.09 & -1930.1 & 8592.7 \\
(mb)           &$\pm$ 0.17 &$\pm$ 0.63 &$\pm$ 0.50 &$\pm$ 1.68 &$\pm$ 2.64 &$\pm$ 4.84 & $\pm$ 749.8 &$\pm$ 931.1 \\
\hline          
$\beta$      & 4.37  & 1.95  & 5.33 & 1.73  & 0.03    & 33.60 & 0.79      & 7.33 \\
(GeV)$^{-2}$ &$\pm$ 0.05 &$\pm$ 0.07 &$\pm$ 1.60 &$\pm$ 0.14 &$\pm$ 4.21 &$\pm$ 41.74 & $\pm$ 0.14 &$\pm$ 0.15 \\
\hline    
$\alpha'$   &          &          & 0.57 &  &  &  &  &  \\
(GeV)$^{-2}$ & & &$\pm$ 0.14 & & & &  &\\
\hline
\end{tabular}
\caption{Set of best-fit parameters for the case with the Odderon.}
\end{table}

As it can be seen from Table 1, only the $b_1^-,\ \alpha_{RP}^-(0),\ C_{RP}^+,\ \beta_O,\ \beta_R^+$ and $\beta_R^-$ parameters (6 out of 35) are not well determined (more than 15\% error).

The resulting value of $\chi^2$ is
\begin{equation}
\label{eq:46}
\chi^2_{dof}=2.46\ ,
\end{equation}
an excellent value if we consider the fact that we did not take into account the systematic errors of the experimental data.

The partial value of $\chi^2$, corresponding only to $t=0$ ($\sigma_T$ and $\rho$) data is 
\begin{equation}
\label{eq:47}
\left.\chi^2_{dof}\right\vert_{t=0}=1.42\ ,
\end{equation}
an acceptable value (276 experimental forward points took into account).
Of course, better $\chi^2$ values can be obtained in fitting \textit{only} the $t=0$ data, as it is in often made in phenomenological papers.
However, it is obvious that, in a global fit including non-forward data, the corresponding $t=0$ parameters will be modified and therefore a higher $\chi^2$ value will be obtained.
The $t=0$ and $t\neq 0$ data are certainly independent but the parameter values are obviously correlated in a global fit.

We plot our fit and predictions for $d\sigma/dt$ data at $\sqrt{s}=52.8$ GeV (Fig.~2), at the RHIC energy values $\sqrt{s}=200$ GeV and $\sqrt{s}=500$ GeV  (Figs.~3 and 4), at the commissioning run energy value $\sqrt{s}=900$ GeV which will be performed in November-December 2007 at LHC (Fig.~5) at the Tevatron energy $\sqrt{s}=1.96$ TeV (Fig.~6) and at the LHC energy value $\sqrt{s}=14$ TeV (Fig.~7).

By comparing Figs. 1 and 2 we can see the huge improvement induced by the Odderon in describing the difference between $pp$ and $\bar pp$ differential cross-sections at $\sqrt{s}=52.8$.
This difference fixes, in fact, as precisely as possible, the magnitude of the Odderon contribution.
The description of the data at $\sqrt{s}=52.8$ GeV as offered by our approach is the best one existing in literature.

It has to ne noted that the structure (dip) region moves slowly, with increasing energy, from $\vert t\vert\approx 1.35$ GeV$^2$ at $\sqrt{s}=52.8$ GeV towards
$\vert t\vert \simeq 0.35$ GeV$^2$ at $\sqrt{s}=14$ TeV.

As it can be already noticed in Figs.~2-7, there is a difference between the $pp$ and $\bar pp$ differential cross-sections. 
This difference is more clearly exposed in Figs.~8-13 where we plot at the same energies as in Figs.~2-7, the quantity
\begin{equation}
\label{eq:48}
\Delta\left(
\frac{d\sigma}{dt}\right)(s,t)\equiv
\left\vert
\left(\frac{d\sigma}{dt}\right)^{\bar pp}(s,t)-
\left(\frac{d\sigma}{dt}\right)^{pp}(s,t)
\right\vert\ .
\end{equation}
There is an interesting phenomenon of oscillations present in $\Delta(\frac{d\sigma}{dt})$, due of the oscillations present in the Heisenberg-type amplitude $F_+^H(s,t)$ and in the maximal Odderon amplitude $F_-^{MO}(s,t)$.
Unfortunately, we can not directly test the existence of these oscillations at RHIC and LHC energies, simply because we will not have both $pp$ and $\bar pp$ accelerators at these energies.
However a chance to detect these oscillations at the RHIC energy $\sqrt{s}=500$ GeV still exists, simply because the UA4/2 Collaboration already performed a high-precision $\bar pp$ experiment at a very close energy - 541 GeV (see Fig.~14).
By performing a very precise experiment at the RHIC energy $\sqrt{s}=500$ GeV and by combining the corresponding $pp$ data with the UA4/2 $\bar pp$ high-precision data one has a non-negligible chance to detect an oscillation centered around $\vert t\vert\simeq 0.9$ GeV$^2$ and therefore to detect the Odderon.
The oscillation centered around $\vert t\vert\simeq 0.15$ GeV$^2$ is slightly more ambiguous because it involves also reggeon contributions.
It is precisely the oscillation centered around $\vert t\vert\simeq 0.9$ GeV$^2$
 which is the reminder of the already seen oscillation centered around $\vert t\vert\simeq 1.35$ GeV$^2$ at the ISR energy $\sqrt{s}=52.8$ GeV.
 
 Of course, it will be desirable to perform also a fixed target experiment at RHIC in order to measure the difference of the total cross-sections $\Delta\sigma(s)$ (see eq.~(\ref{eq:2})) at
 \begin{equation}
\label{eq:49}
50\leqslant \sqrt{s}\leqslant 500 \mbox{ GeV}\ .
\end{equation}
A conclusive proof of the existence of the Odderon would be to establish that $\Delta\sigma\neq 0$ at these energies.
Moreover, the maximal Odderon induces a spectacular effect, because of the phenomenological sign of the $O_1$ parameter: at $\sqrt{s}=500$ GeV, where the contribution of the odd-signature secondary reggeons will be negligible, the $pp$ total cross-section will be higher than the $\bar pp$ total cross-section.
Unfortunately, due to the smallness of the $O_1$ parameter and to the slow $\ln s$-physics, $\Delta\sigma(\sqrt{s}=500$ GeV) would be only of the order of -1.15 mb, which might be impossible to establish experimentally in a non-ambiguous way, due to the experimental errors in $\sigma_T$.
The difference $\Delta\sigma(s)$ would reach the relatively small value -33.5 mb at $\sqrt{s}=10^{19}$ GeV.

There are other possibilities to look for the Odderon at  RHIC, like the measurement of $d\sigma/dt$ in the very small $t$-range
\begin{equation}
\label{eq:50}
0.003\leqslant\vert t\vert \leqslant 0.04\mbox{ GeV}^2\ ,
\end{equation}
in order to extract the $\rho$-parameter, or the measurement of $A_{NN}$.

The maximal Odderon predictions for $t=0$ observables are:
\begin{alignat}{1}
&\sigma_T^{pp}(\sqrt{s}=500 \mbox{ GeV})=62.8\mbox{ mb}\ ,\label{eq:51}\\
&\Delta\sigma(\sqrt{s}=500 \mbox{ GeV})=-1.15\mbox{ mb}\ ,\label{eq:52}\\
&\rho_{pp}(\sqrt{s}=500 \mbox{ GeV})=0.154\ ,\label{eq:53}\\
&\Delta\rho(\sqrt{s}=500 \mbox{ GeV},\ t=0)=0.004\ .\label{eq:54}
\end{alignat}
The participants at the workshop "Odderon Searches at RHIC", hold at BNL in September 2005, concluded that the best available setup for the experimental search for the Odderon is the proposed combination of STAR experiment and Roman pots at $pp2pp$ experiment,  described in the proposal "Physics with Tagged Forward Protons with the STAR detector at RHIC".
They also concluded that the most unambiguous signature of the Odderon is to detect a non-zero difference between $pp$ and $\bar pp$ differential cross-sections  at $\sqrt{s}=500$ GeV, as described above.
RHIC is an ideal place for discovering the Odderon and therefore testing QCD and CGC~\cite{Guryn:yk}.

LHC is also a good place to discover the Odderon.
We predict
\begin{alignat}{1}
&\sigma_T^{pp}(\sqrt{s}=14 \mbox{ TeV})=123.32\mbox{ mb}\ ,\label{eq:55}\\
&\Delta\sigma(\sqrt{s}=14 \mbox{ TeV})=-3.92\mbox{ mb}\ ,\label{eq:56}\\
&\rho_{pp}(\sqrt{s}=14 \mbox{ TeV},\  t=0)=0.103\label{eq:57}\ ,
\end{alignat}
and
\begin{equation}
\label{eq:58}
\Delta\rho(\sqrt{s}=14 \mbox{ TeV},\ t=0)=0.094\ .
\end{equation}
A $\rho^{pp}$-measurement at LHC would be certainly a very important test of the maximal Odderon, given the fact that our prediction is sufficiently lower than what dispersion relations without Odderon contributions could predict ($\rho\simeq 0.12 - 0.14$).

However, one must take into account the ambiguities related to the extraction of $\rho$, which is a semi-theoretical parameter.
Such ambiguities could be avoided if the ATLAS experiment, which will perform a dedicated small-$t$ experiment at LHC~\cite{Efthymio:2005bl}, would apply our new method for the determination of the real part of the hadron elastic scattering at small angles and high energies~\cite{Gauron:2004ba}.
This method provides a strong constraint on the parameter $\rho$ and is therefore crucial in detecting new phenomena in the Standard Model (like the Odderon presence) or even signatures of new physics (e.g., violation of dispersion relations~\cite{Bourrely:2005qh}, which would lead to high values of $\rho$ ($\rho\simeq 0.21$).

There are several other proposals for detecting the Odderon, summarized in the nice review written by Ewerz~\cite{Ewerz:2003xi}.

%**************************************
\section{Comparison with other models}
%**************************************
Most models of diffraction scattering are constructed so that the crossing-symmetric amplitude $F_+$ dominates at high energies for all $t$~ 
\cite{Bourrely:1984fz,Cheng:1970bi,Cheng:1972ik,Chou:1979pz,Chou:1983zi}.
The contributions to $F_-$ are usually Regge-like and consequently have largely disappeared by ISR energies.
Hence these models predict equality of $(d\sigma/dt)^{\bar pp}$ and $(d\sigma/dt)^{pp}$, in serious contradiction with the ISR data at $\sqrt{s}=52.8$ GeV.
An an example, one could contemplate Fig.~6 of Ref.~\cite{Bourrely:2002wr}, in order to see the big discrepancy between the Bourrely, Soffer and Wu model and the $\bar pp$ data at $\sqrt{s}=52.8$ GeV in the critical structure region centered around $\vert t\vert=1.35$ GeV$^2$.

There are several models in the literature which include a crossing-odd amplitude $F_-$ that remains important at ISR energies.

Between them, the most similar in spirit, as compared with our own approach, is the model of Donnachie and Landshoff~\cite{Donnachie:1984xq}.
First, Donnachie and Landshoff include an Odderon contribution, described as the exchange of 3-(nonreggeized) gluons, calculated in pQCD.
Second, these authors include the Regge poles and cuts contributions as an important component of their amplitudes.

One has to note that their $F_-$ dominates at large $t$ for sufficiently high energies, as it does for us.
However their $F_-$ becomes constant at fixed $t$ as $s\to\infty$ and, according to the authors, may well begin to decrease when higher order corrections are taken into account.
In contrast our maximal Odderon $F_-^{MO}$ grows at fixed $t$, in a certain $t$ range beyond the dip-shoulder region.

On a strictly theoretical level, there is no reason, as explained above, to apply pQCD in a moderate-$t$ region.
On a phenomenological level, one has to note that the Donnachie-Landshoff model, as it can be seen from Fig.~15 (as given in \cite{Ewerz:2003xi}), do not describe well the $pp$ and $\bar pp$ data at $\sqrt{s}=52.8$ GeV in the critical structure region centered around $\vert t\vert=1.35$ GeV$^2$.
We therefore think that the existing data favor the maximal Odderon as compared with the Donnachie-Landshoff 3-gluon Odderon.

Another interesting Odderon model was formulated by Islam \textit{et al.}~\cite{Islam:1983bt,Islam:1987va}, in the framework of the so-called "cloud-core model".
In this model, the nucleon is visualized as a core of valence quarks surrounded by a cloud of quark-antiquark pairs and it is argued that an Odderon amplitude emerges when the cores interact by exchanging a $C=-1\ u\bar u+d\bar d$ state while the clouds undergo maximal diffraction scattering.

The cloud-core model involves therefore an Odderon belonging to the class of the minimal Odderon, which, by itself is unable to account for the already present Odderon effects.
This can be seen from Fig.~16 which shows the prediction of the cloud-core model at the RHIC and CERN-collider energy $\sqrt{s}\simeq 500$ GeV~\cite{Islam:2002au}: a difference between $pp$ and $\bar pp$ differential cross-sections is certainly present in this model but the $\bar pp$ theoretical curve badly miss the experimental points in the region $0.5\lesssim\vert t\vert \lesssim 1.5$ GeV$^2$.

In making theoretical predictions for detecting the Odderon at RHIC and LHC one has to keep always in mind that we have, as an absolute necessity, to provide first an excellent \textit{quantitative} description of the already existing data and especially of the $pp$ and $\bar pp\ d\sigma/dt$ data at $\sqrt{s}=52.8$ GeV.
Otherwise, on the basis of an acceptable qualitative model we can draw wrong quantitative conclusions about an object - the Odderon - which appears, in $pp$ and $\bar pp$ scatterings, as a small correction to the dominant Pomeron contribution, except in particular regions of $s$ and $t$.

%***********************************
\section{Conclusions}
%***********************************
There are very rare cases in the history of physics that a scientific and testable idea is neither proved nor disproved 33 years after its invention.
The Odderon remains an elusive object in spite of intensive research for its experimental evidence.

The main reason for this apparent puzzle is that most of the efforts were concentrated in the study of $pp$ and $\bar pp$ scattering, where the $F_-(s,t)$ amplitude is hidden by the overwhelming $F_+(s,t)$ amplitude.
The most spectacular signature of the Odderon is the predicted difference between $pp$ and $\bar pp$ scattering at high $s$ and relatively small $t$.
However, it happens that, after the closure of ISR, which offered the first strong hint for the existence of the Odderon, there is no place in the world where $pp$ and $\bar pp$ scattering are or will be measured at the same time. 
This is the main reason of the non-observation till now of the Odderon.

In this paper, we show that we can escape from this unpleasant situation by performing a high-precision measurement of $d\sigma/dt$ at RHIC, at $\sqrt{s}= 500$ GeV, and by combining these future data with the already present high-precision UA4/2 data at $\sqrt{s}= 541$ GeV.

There is no doubt about the theoretical evidence for the Odderon both in QCD and CGC.
The Odderon is a fundamental object of these two approaches and it has to be found at RHIC and LHC if QCD and CGC are right.

\section*{Acknowledgments}
One of us (R.A.) thanks FAPESP (Brazil) for the doctoral grant No. 03/00228-0 and 04/10619-9 for France and Prof. Pascal Debu for the kind hospitality at LPNHE Paris, where the present work was performed.
The authors are grateful to Professors Claude Bourrely and Evgenij Martynov
 who kindly repeated our calculations. This crosscheck proved to be very
useful.
The authors also thank Mlle Josette Durin for the kind technical assistance in the preparation of the present paper.

%\bibliographystyle{h-physrev}
%\bibliography{Odderon}

\vspace{-15cm}

\begin{figure}[h]
\begin{center}
\includegraphics[scale=0.9]{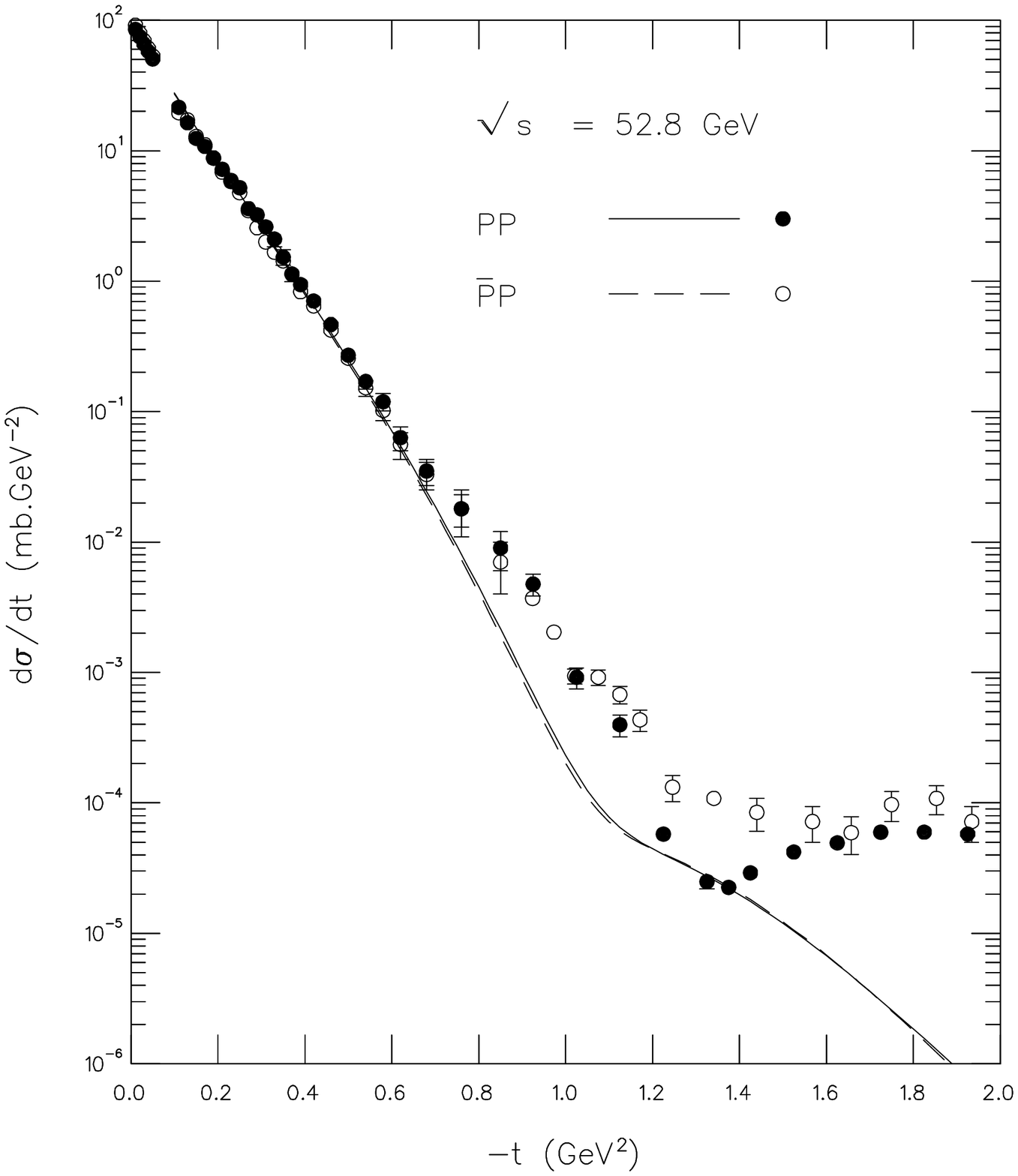}
\caption{$pp$ and $\bar pp\ d\sigma/dt$ predictions for the case without the Odderon, together with the experimental points, at $\sqrt{s}=52.8$ GeV.}
\label{fig:1}
\end{center}
\end{figure}

\begin{center}
\begin{figure}[ht]
\includegraphics[scale=0.9]{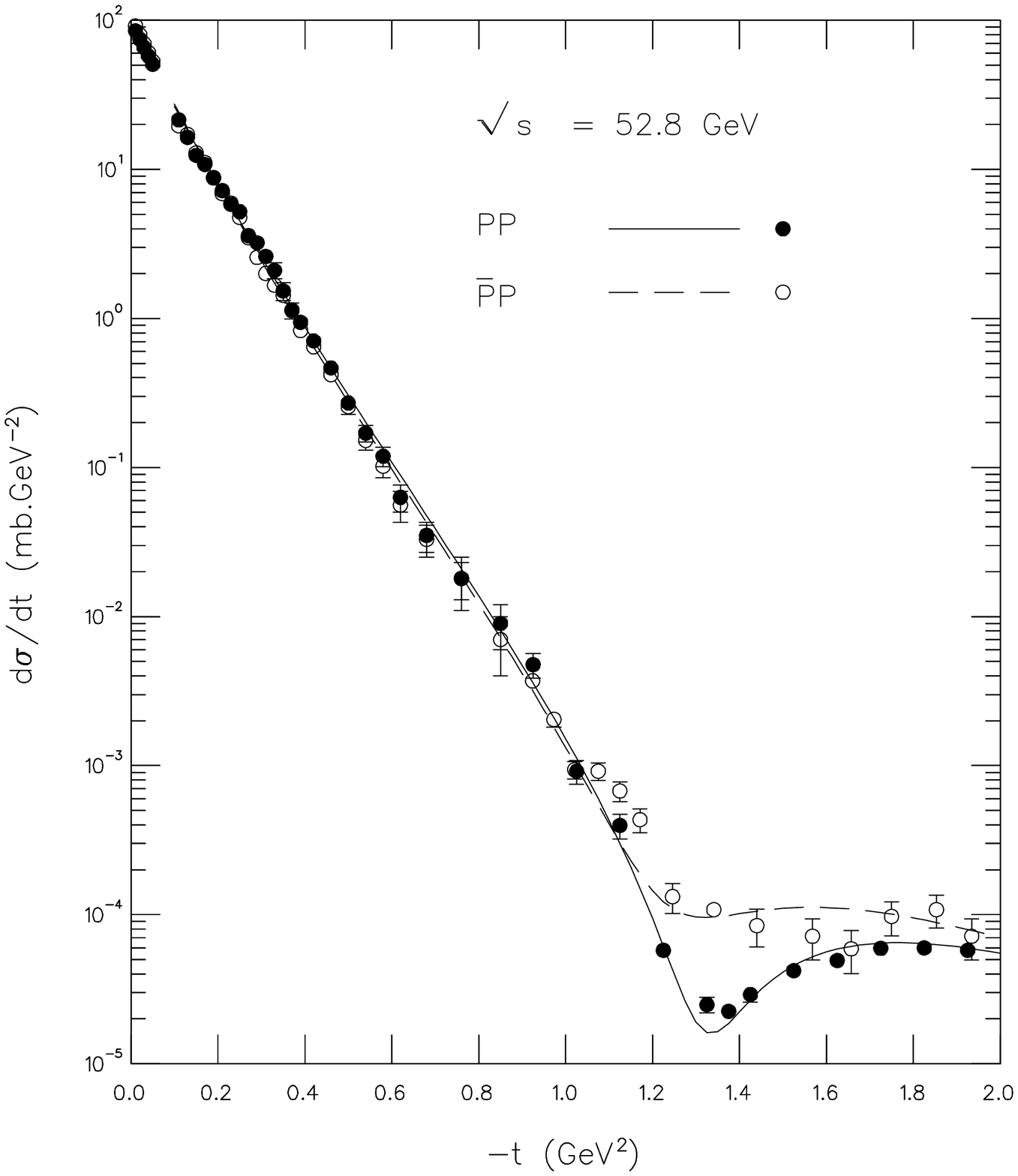}
\caption{$pp$ and $\bar pp\ d\sigma/dt$ predictions for the case with the Odderon, together with the experimental points, at $\sqrt{s}=52.8$ GeV.}
\label{fig:2}
\end{figure}
\end{center}

\begin{center}
\begin{figure}[ht]
\includegraphics[scale=0.9]{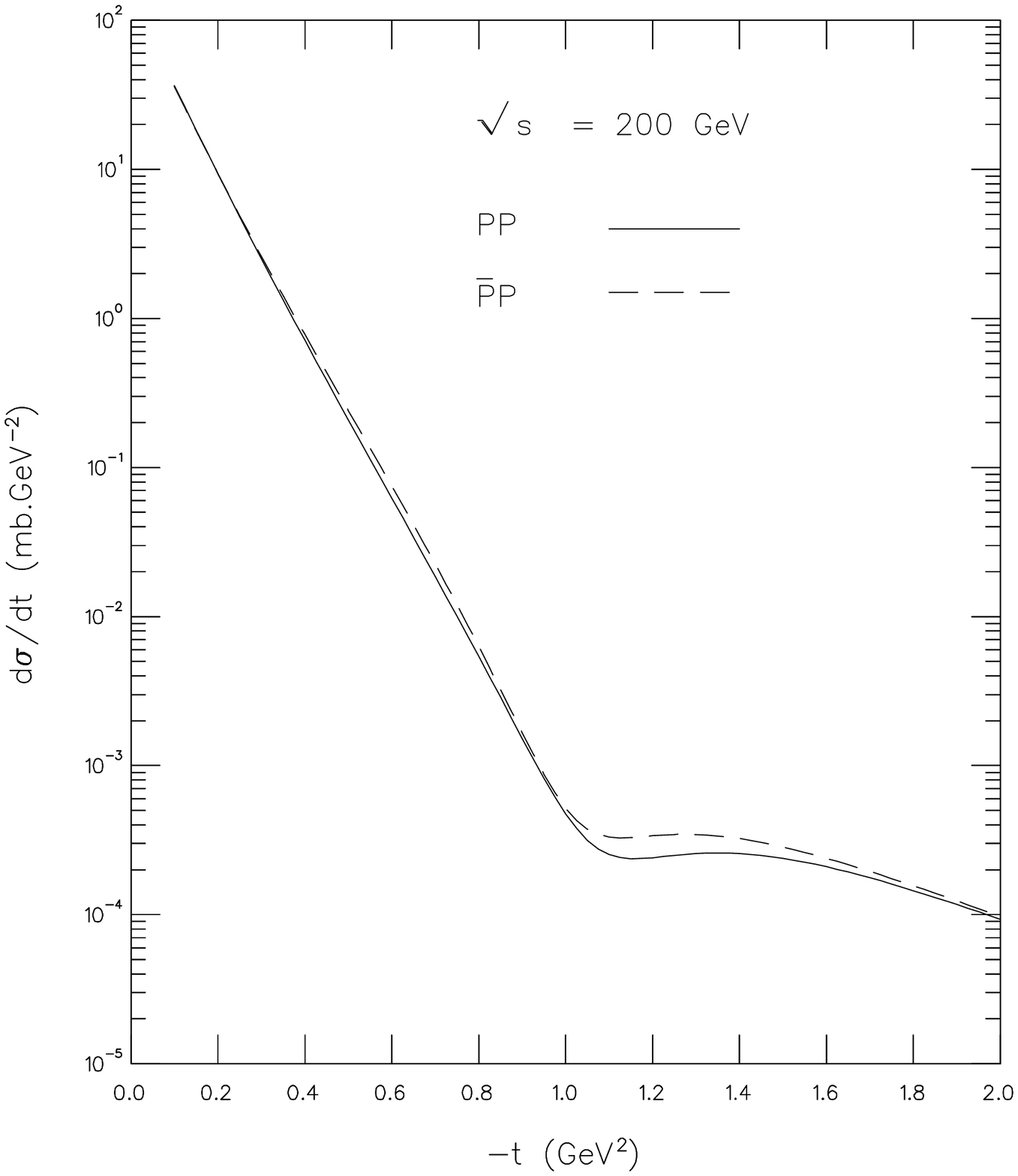}
\caption{$pp$ and $\bar pp\ d\sigma/dt$ predictions for the case with the Odderon, at 
$\sqrt{s}=200$ GeV.}
\label{fig:3}
\end{figure}
\end{center}

\begin{center}
\begin{figure}[ht]
\includegraphics[scale=0.9]{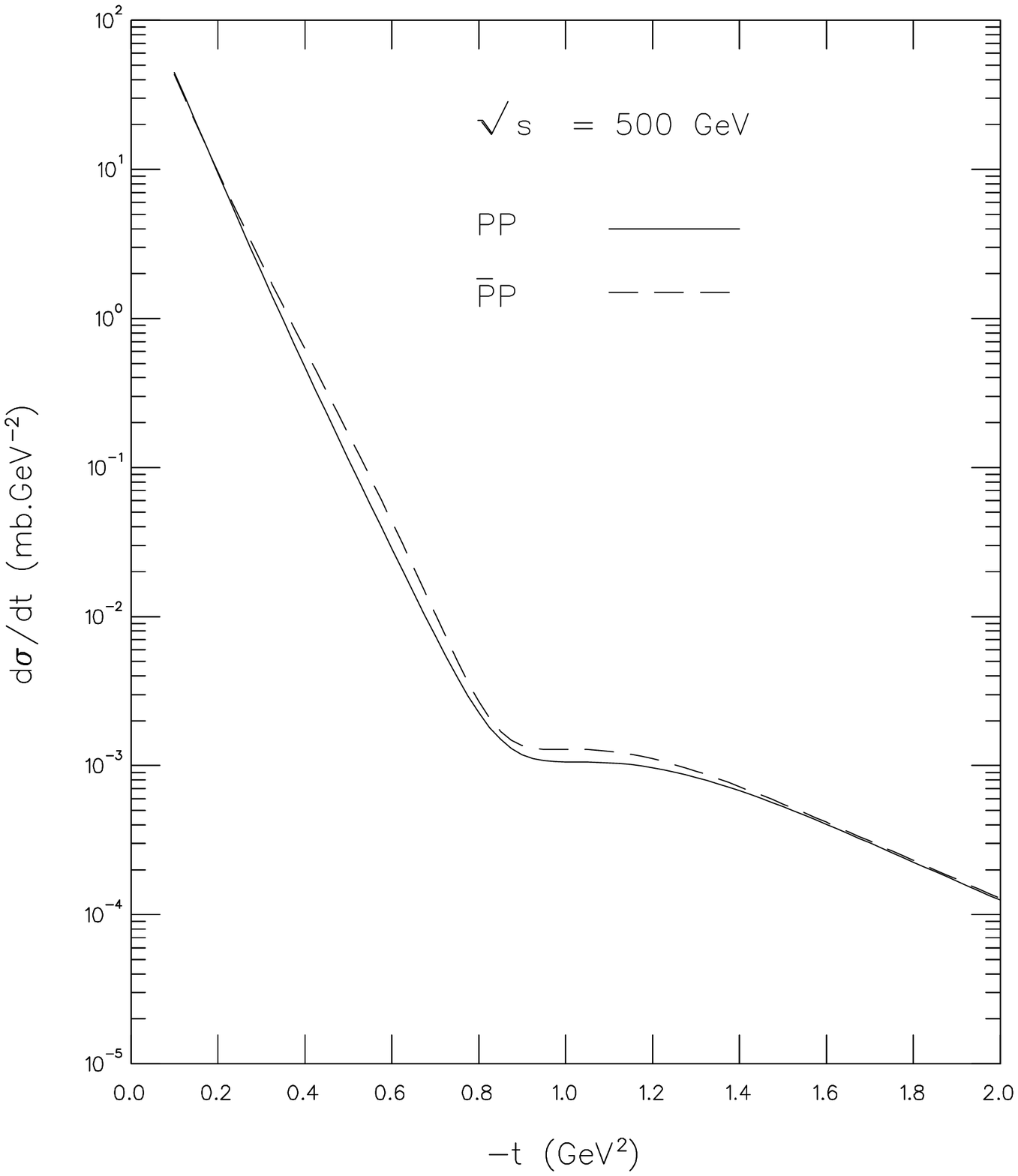}
\caption{$pp$ and $\bar pp\ d\sigma/dt$ predictions for the case with the Odderon, at 
$\sqrt{s}=500$ GeV.}
\label{fig:4}
\end{figure}
\end{center}

\begin{center}
\begin{figure}[ht]
\includegraphics[scale=0.9]{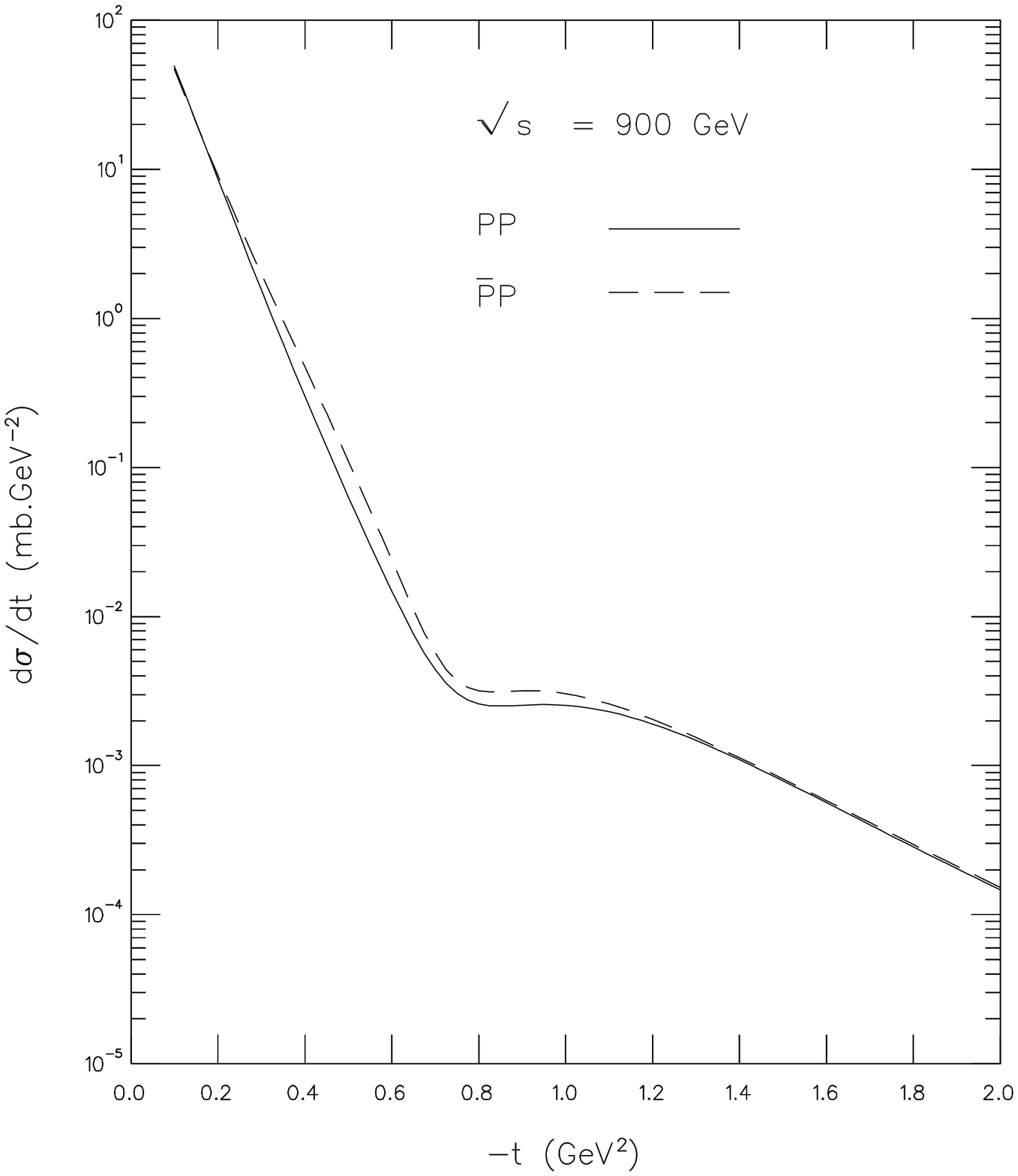}
\caption{$pp$ and $\bar pp\ d\sigma/dt$ predictions for the case with the Odderon, at 
$\sqrt{s}=900$ GeV.}
\label{fig:5}
\end{figure}
\end{center}

\begin{center}
\begin{figure}[ht]
\includegraphics[scale=0.9]{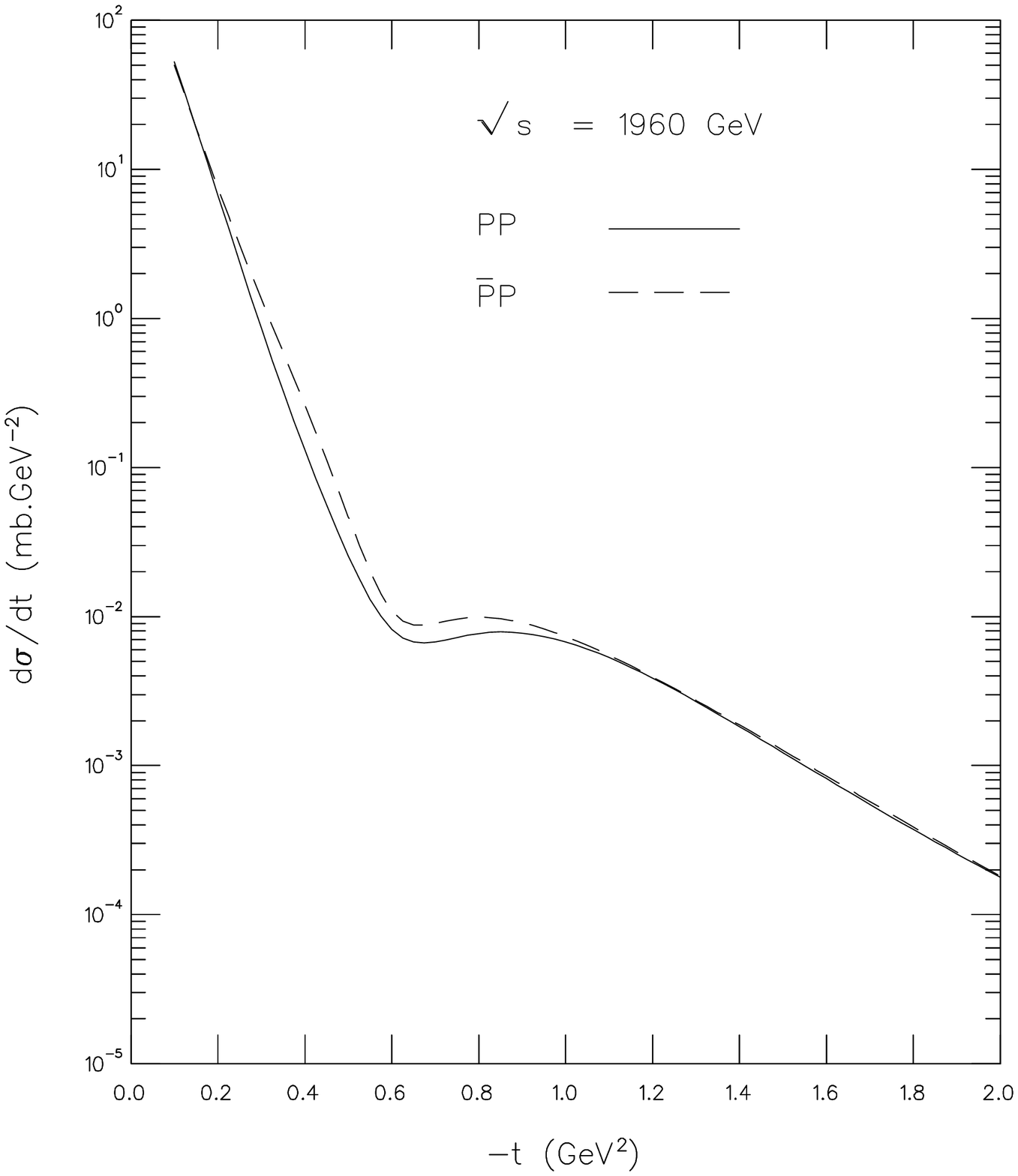}
\caption{$pp$ and $\bar pp\ d\sigma/dt$ predictions for the case with the Odderon, at 
$\sqrt{s}=1960$ GeV.}
\label{fig:6}
\end{figure}
\end{center}

\begin{center}
\begin{figure}[ht]
\includegraphics[scale=0.9]{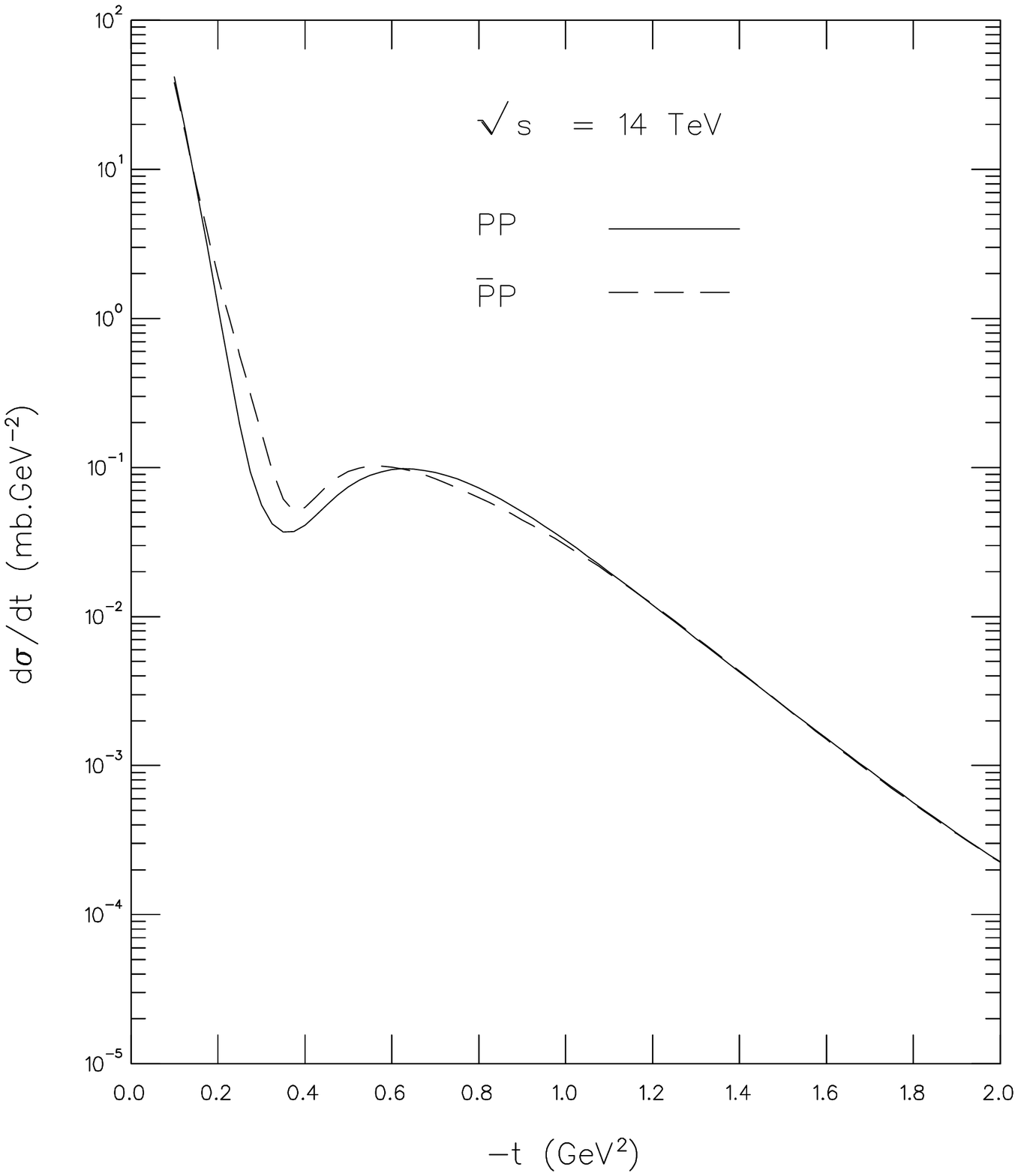}
\caption{$pp$ and $\bar pp\ d\sigma/dt$ predictions for the case with the Odderon, at 
$\sqrt{s}=14$ TeV.}
\label{fig:7}
\end{figure}
\end{center}

\begin{center}
\begin{figure}[ht]
\includegraphics[scale=0.9]{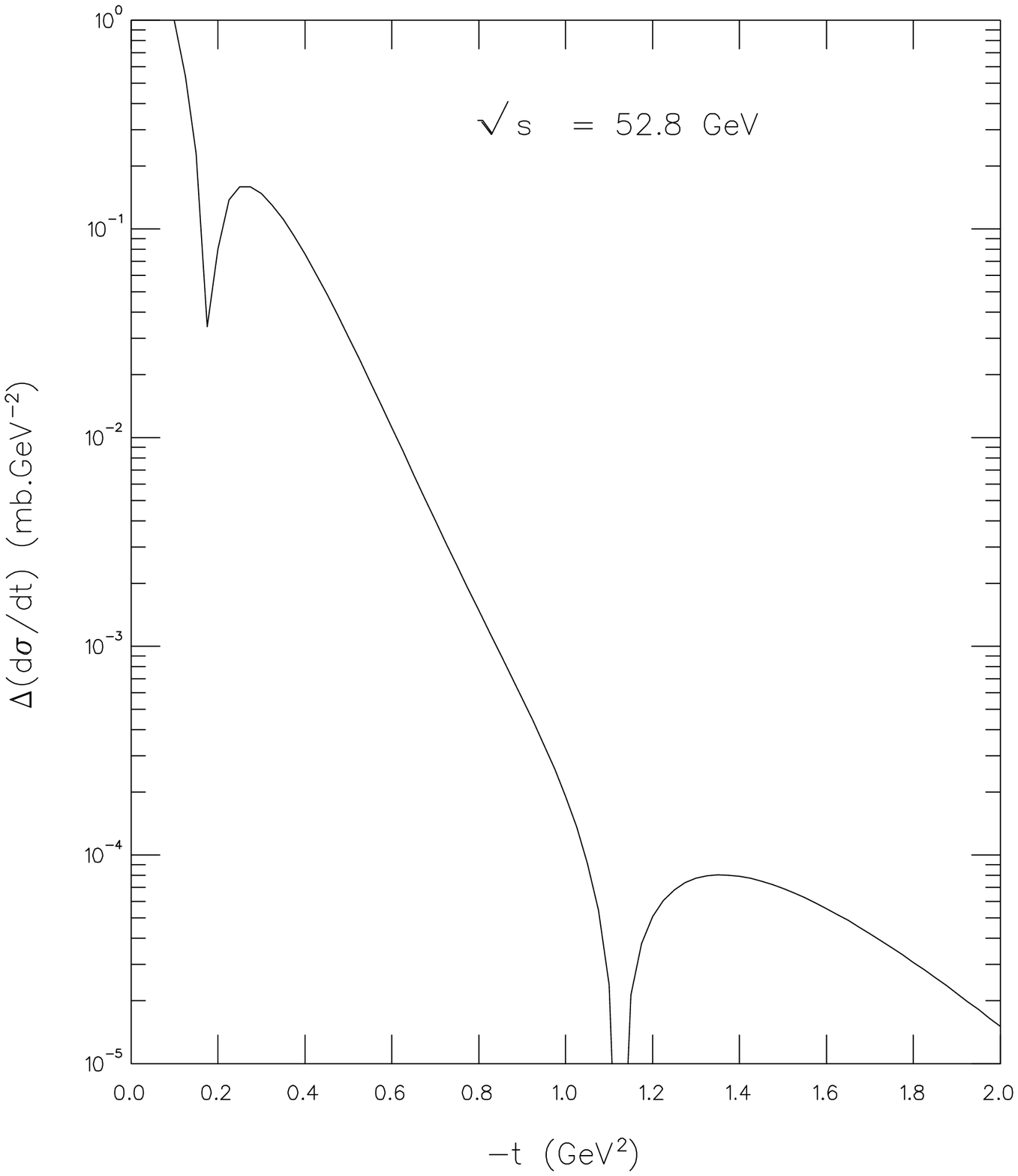}
\caption{Prediction for $\Delta(d\sigma/dt)$ (eq.~\protect\ref{eq:48}), at 
$\sqrt{s}=52.8$ GeV.}
\label{fig:8}
\end{figure}
\end{center}

\begin{center}
\begin{figure}[ht]
\includegraphics[scale=0.9]{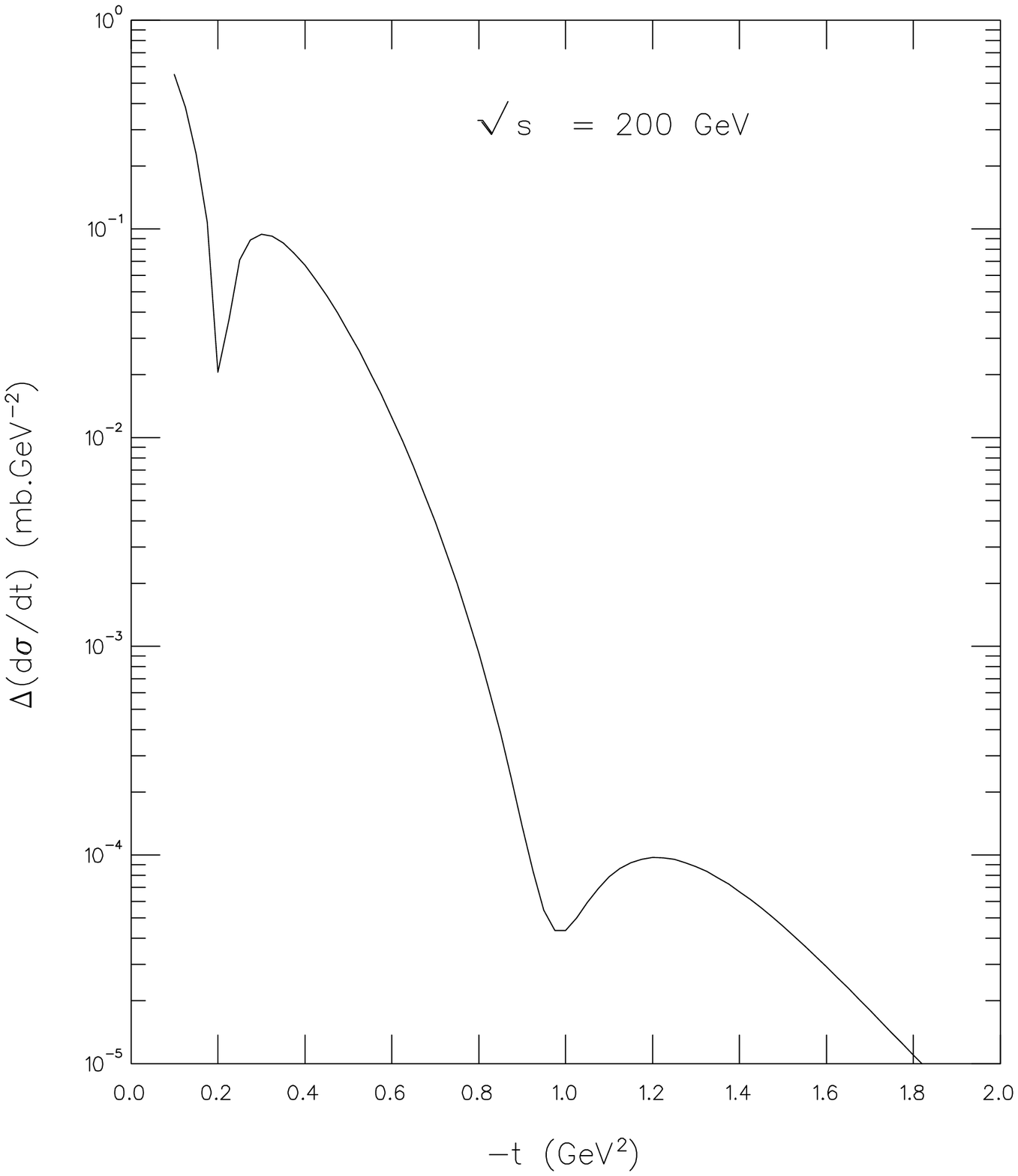}
\caption{Prediction for $\Delta(d\sigma/dt)$ (eq.~\protect\ref{eq:48}), at 
$\sqrt{s}=200$ GeV.}
\label{fig:9}
\end{figure}
\end{center}

\begin{center}
\begin{figure}[ht]
\includegraphics[scale=0.9]{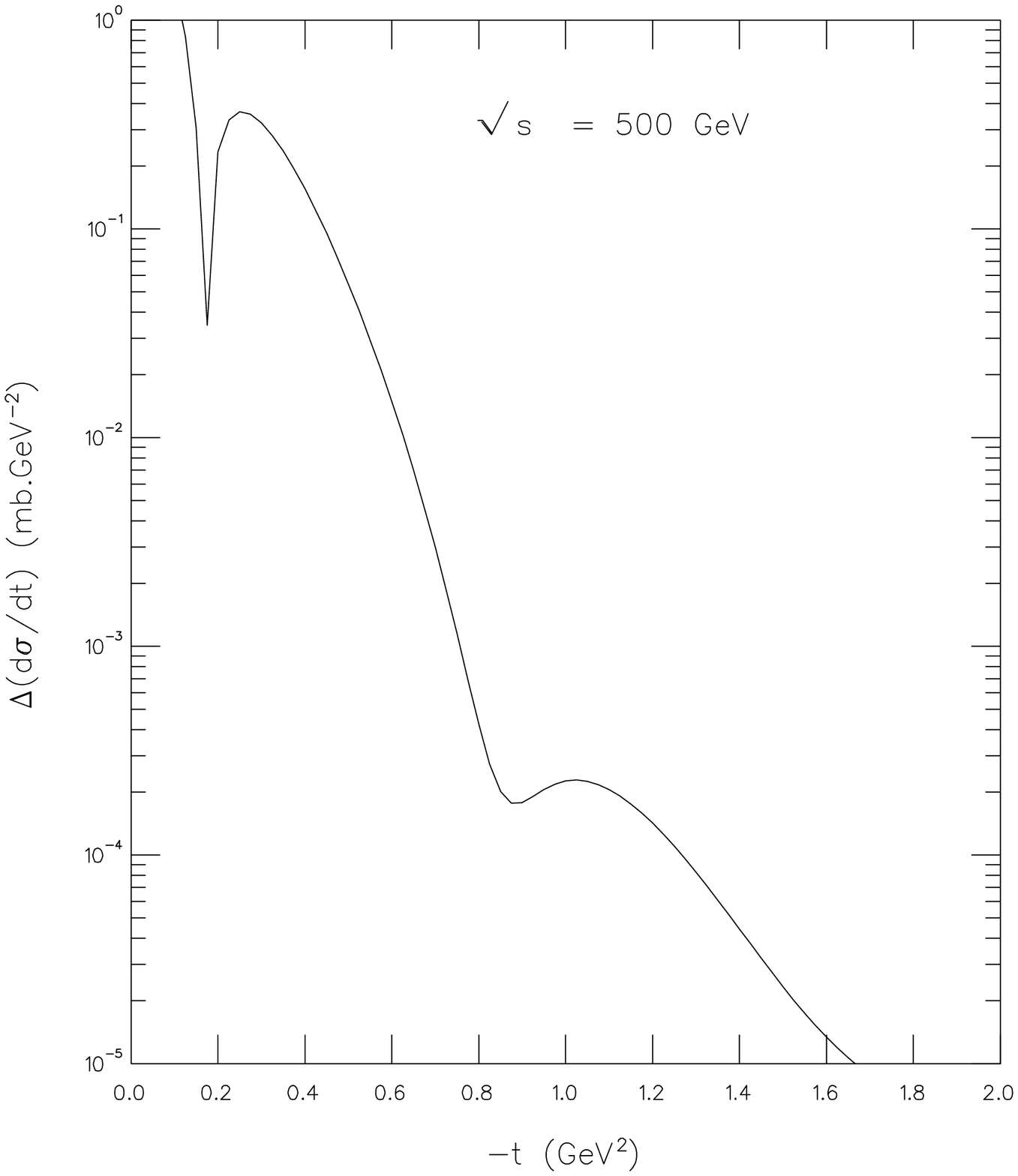}
\caption{Prediction for $\Delta(d\sigma/dt)$ (eq.~\protect\ref{eq:48}), at 
$\sqrt{s}=500$ GeV.}
\label{fig:10}
\end{figure}
\end{center}

\begin{center}
\begin{figure}[ht]
\includegraphics[scale=0.9]{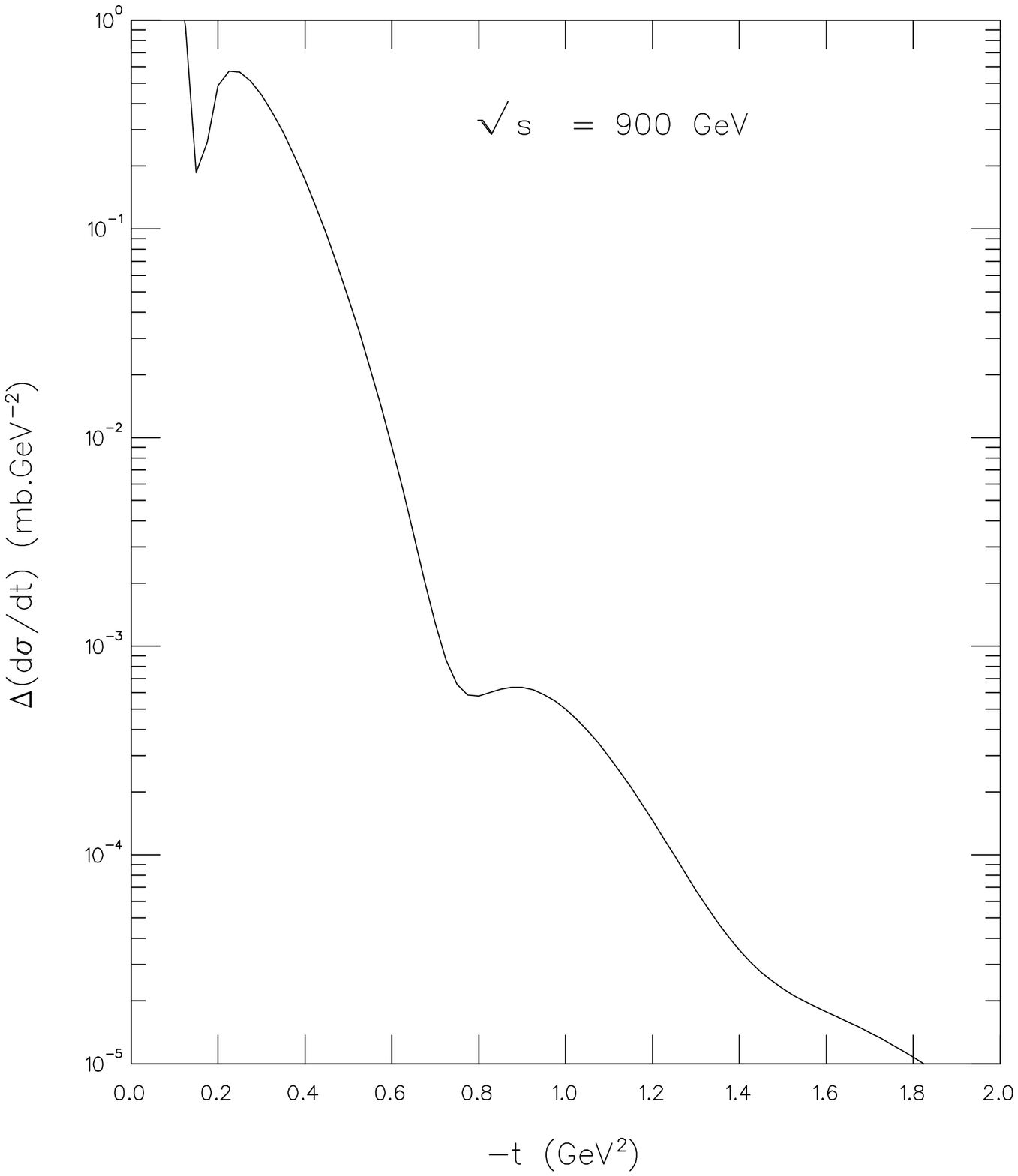}
\caption{Prediction for $\Delta(d\sigma/dt)$ (eq.~\protect\ref{eq:48}), at 
$\sqrt{s}=900$ GeV.}
\label{fig:11}
\end{figure}
\end{center}

\begin{center}
\begin{figure}[ht]
\includegraphics[scale=0.9]{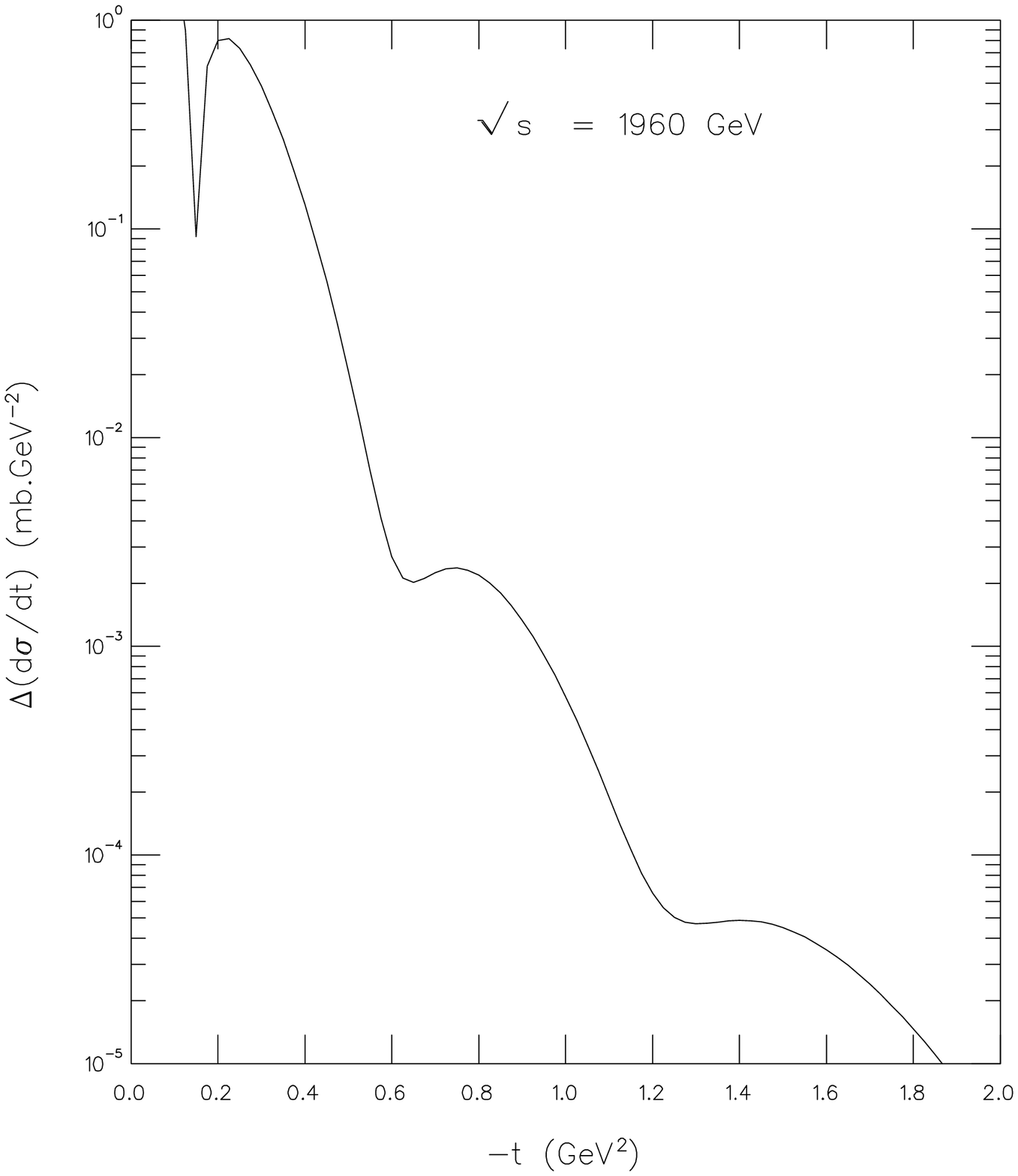}
\caption{Prediction for $\Delta(d\sigma/dt)$ (eq.~\protect\ref{eq:48}), at 
$\sqrt{s}=1960$ GeV.}
\label{fig:12}
\end{figure}
\end{center}

\begin{center}
\begin{figure}[ht]
\includegraphics[scale=0.9]{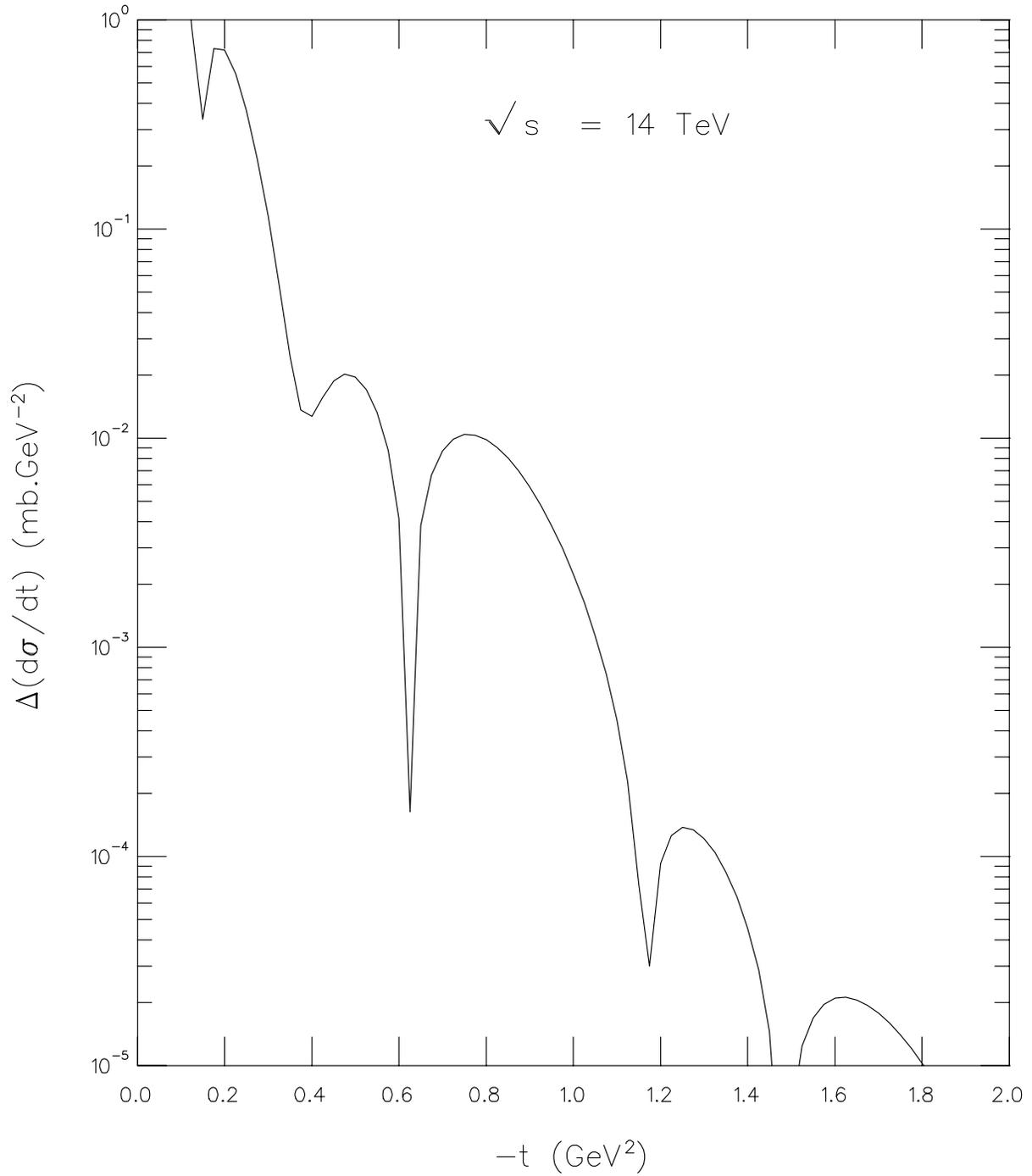}
\caption{Prediction for $\Delta(d\sigma/dt)$ (eq.~\protect\ref{eq:48}), at 
$\sqrt{s}=14$ TeV.}
\label{fig:13}
\end{figure}
\end{center}

\begin{center}
\begin{figure}[ht]
\includegraphics[scale=0.9]{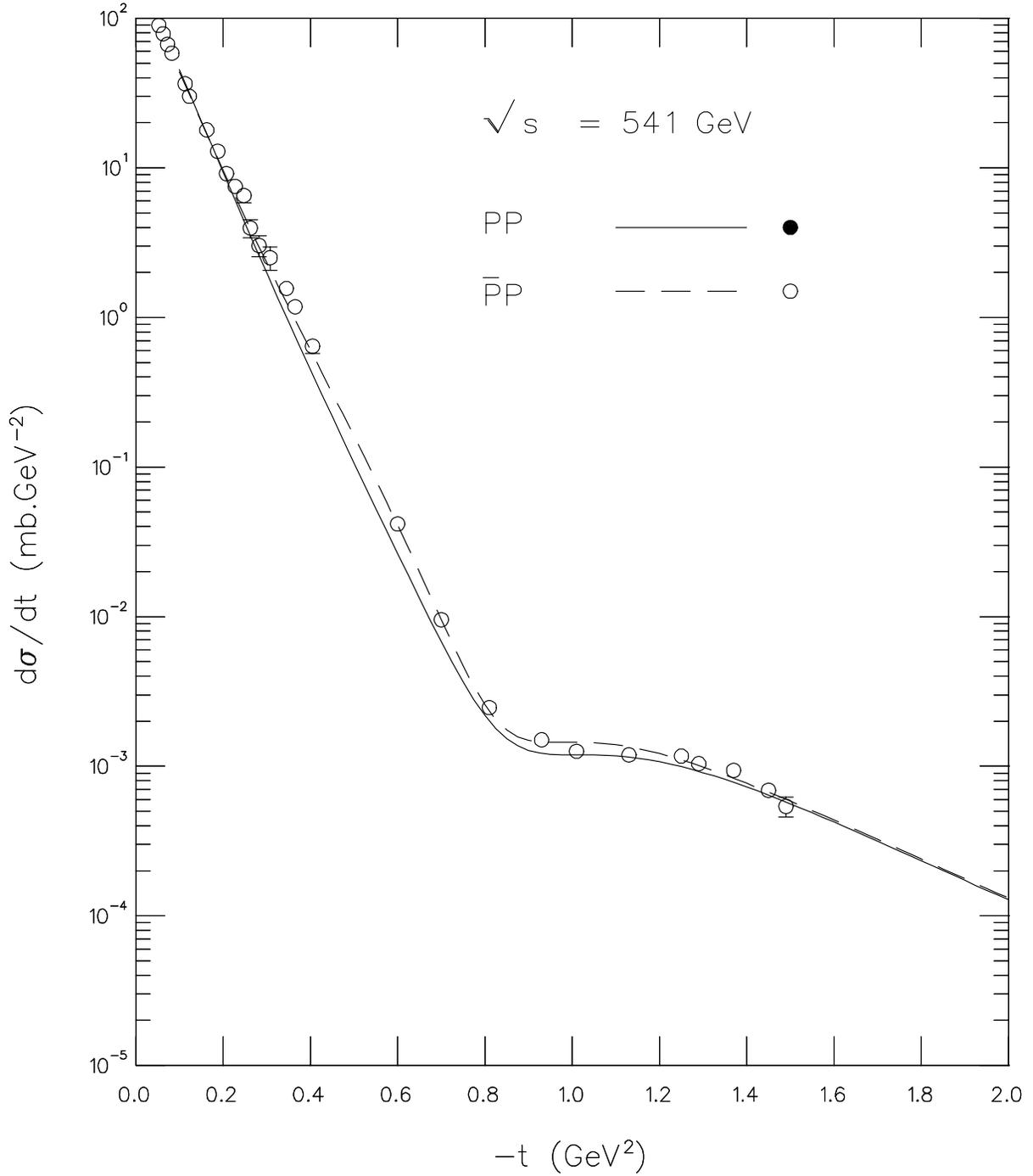}
\caption{$pp$ and $\bar pp\ d\sigma/dt$ predictions for the case with the Odderon, together with the UA4/2 $\bar pp$ experimental points, at $\sqrt{s}=541$ GeV. }
\label{fig:14}
\end{figure}
\end{center}

\begin{figure}[t]
\begin{center}
\includegraphics[scale=0.8]{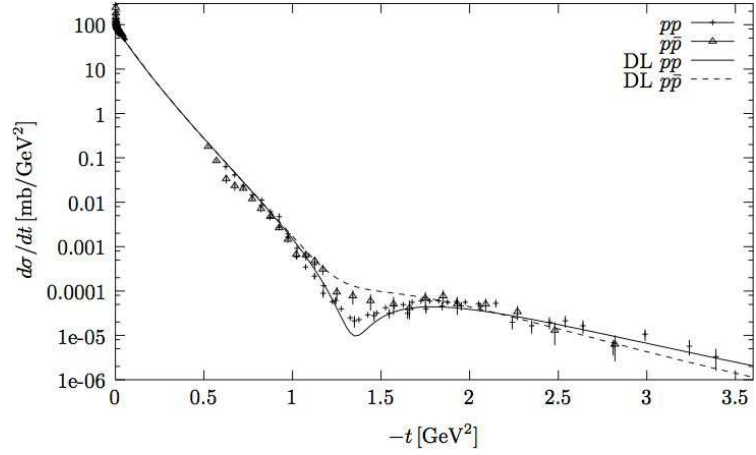}
\caption{Differential cross section for elastic $pp$ and $p\bar{p}$
for $\sqrt{s}=52.8$ GeV together with the Donnachie--Landshoff (DL)
fit \protect\cite{Donnachie:1984xq}. This figure is taken from Ref.\protect\cite{Ewerz:2003xi}, where it appears as Fig. 10.}
\label{fig:15}
\end{center}
\end{figure}

\begin{figure}[b]
\begin{center}
\includegraphics[scale=0.6]{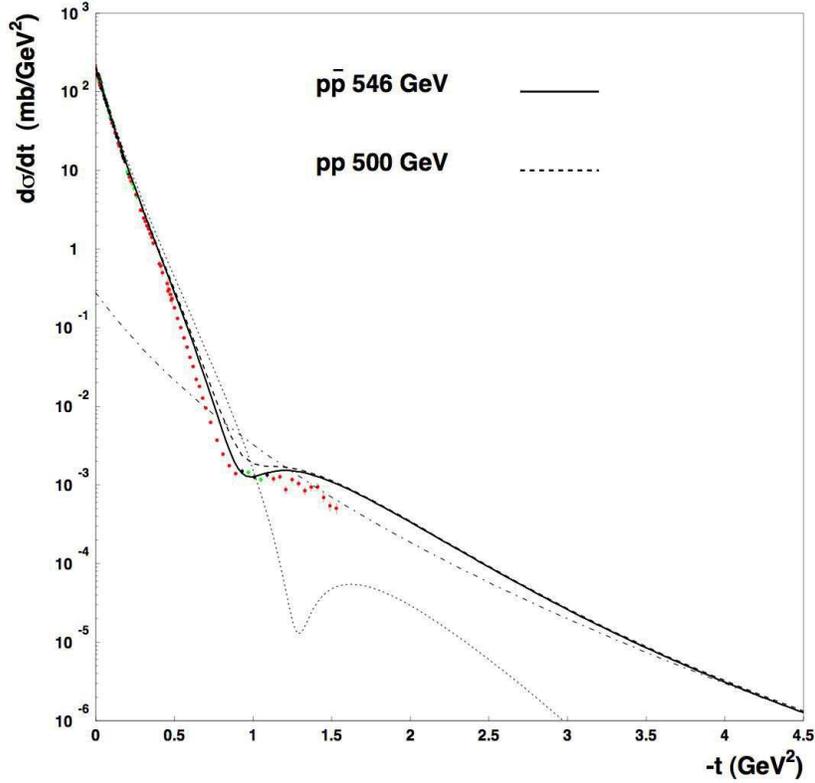}
\caption{Figure 3 from Ref.\protect\cite{Islam:2002au}. The solid curve represents the calculated $d\sigma/dt$ for $\bar pp$ at $\sqrt{s}=546$ GeV and the thick-dashed curve shows the predicted $d\sigma/dt$ for $pp$ at $\sqrt{s}=500$ GeV. }
\label{fig:16}
\end{center}
\end{figure}

\end{document}